\documentclass[showpacs,10pt,twocolumn,prb]{revtex4-1}

\usepackage{amsmath}
\usepackage{amssymb}
\usepackage{graphicx}
\usepackage{amssymb}
\usepackage{graphics}
\usepackage{epsfig}
\usepackage{color}

\begin{document}

\title{Multiband behavior and non-metallic low-temperature state of K$_{0.50}$Na$_{0.24}$Fe$_{1.52}$Se$_{2}$}
\author{Hyejin Ryu,$^{1,2,\ddag}$ F. Wolff-Fabris,$^{3\dag}$ J. B. Warren,$^{4}$ M. Uhlarz,$^{3}$ J. Wosnitza,$^{3,5}$ and C. Petrovic$^{1,2,\ddag}$}
\affiliation{$^{1}$Condensed Matter Physics and Materials Science Department, Brookhaven
National Laboratory, Upton, New York 11973, USA\\
$^{2}$Department of Physics and Astronomy, Stony Brook University, Stony Brook, New York 11794-3800, USA\\
$^{3}$Hochfeld-Magnetlabor Dresden (HLD), Helmholtz-Zentrum Dresden-Rossendorf, D-01314 Dresden, Germany\\
$^{4}$Instrument Division, Brookhaven National Laboratory, Upton, New York 11973, USA\\
$^{5}$Institut f\"{u}r Festk\"{o}rperphysik, TU Dresden, D-01062, Dresden, Germany}
\date{\today}

\date{\today}

\begin{abstract}

We report evidence for multiband transport and an insulating
low-temperature normal state in superconducting K$_{0.50}$Na$_{0.24}$Fe$_{1.52}$Se$_{2}$
with $T_{c}\approx 20$ K. The temperature-dependent
upper critical field, $H_{c2}$, is well described by a two-band BCS model.
The normal-state resistance, accessible at low
temperatures only in pulsed magnetic fields, shows an insulating logarithmic
temperature dependence as $T \rightarrow 0$ after superconductivity is suppressed.
This is similar as for high-$T_{c}$ copper oxides and granular type-I superconductors,
suggesting that the superconductor-insulator transition observed in high
magnetic fields is related to intrinsic nanoscale phase separation.

\end{abstract}

\pacs{72.80.Ga, 74.25.F-, 74.81.-g, 74.70.Xa}
\maketitle

After the discovery of LaFeAsO$_{1-x}$F$_{x}$ with $T_c=26$ K\cite{Kamihara}
many efforts have been made to study the temperature dependence
of the upper critical field, $H_{c2}$, of Fe-based superconductors since this
provides valuable insight in the coherence length, anisotropy, electronic
structure, and the pair-breaking mechanism. Binary $\beta$-FeSe and
Fe$_{1+y}$(Te,Se) (FeSe-11 type) as well as arsenic-deficient CuZrSiAs
structure-type superconductors (FeAs-1111 type) feature a Pauli-limited
$H_{c2}$ and are well explained by the single-band Werthamer-Helfand-Hohenberg
(WHH) model.\cite{LeiH,LeiH1,Fuchs,Kida} On the other hand, in most FeAs-1111
type, ternary pnictide (FeAs-122 type), and chalcogenide (FeSe-122 type)
Cu$_{2}$TlSe$_{2}$ Fe-based superconductors $H_{c2}$ can
only be described by two-band models.\cite{Hunte,Jaroszynski,Baily,Mun}
Studies of the normal state below $T_{c}$ in both Cu- and Fe-based high-$T_{c}$ superconductors are rare since very high magnetic
fields are required to suppress the superconductivity. Among the few
exceptions are studies of La$_{2-x}$Sr$_{x}$CuO$_{4}$ and
Bi$_{2}$Sr$_{2-x}$La$_{x}$CuO$_{6}$, where a logarithmic resistivity and
a superconductor-insulator transition (SIT) have been observed in the
normal-state region above $H_{c2}$ and below $T_{c}$.\cite{Ando,Boebinger,Ono}
Similar studies in FeSe-122-type superconductors have
not been available so far due to their air sensitivity and the demanding
experimental conditions of pulsed-field experiments.

In this work, we report on results obtained for
single-crystalline K$_{0.50}$Na$_{0.24}$Fe$_{1.52}$Se$_{2}$ with
$T_{c}\approx20$ K. $H_{c2}(T)$ is well described by a two-band model.
Moreover, when superconductivity is suppressed in high magnetic fields,
the in-plane sample resistance follows $R_{ab} \propto
\ln(T)$ as $T\rightarrow0$, suggesting a SIT, as commonly observed
in granular superconductors.

The K$_{0.50(1)}$Na$_{0.24(4)}$Fe$_{1.52(3)}$Se$_{2.00(5)}$ single crystals
used in this study were synthesized and characterized as described previously
with a nominal composition of starting materials K:Na:Fe:Se = 0.6:0.2:2:2.\cite{Lei00} The as-grown crystals were sealed in a Pyrex tube under vacuum
($\sim$10$^{-1}$ Pa), annealed at 400${^{\circ}}$C for 3 hours, and then
quenched in air in order to increase the superconducting volume fraction.\cite{Lei,Ryu,HanF} Powder x-ray diffraction (XRD) spectra were taken with
Cu $K_{\alpha}$ radiation ($\lambda = 0.15418$ nm) by a Rigaku Miniflex
X-ray machine. The lattice parameters were obtained by refining XRD spectra
using the Rietica software.\cite{Hunter} The elemental analysis was done using a scanning electron microscope (SEM). Magnetization measurements were performed
in a Quantum Design MPMS-XL5. The ac magnetic susceptibility was measured
with an excitation frequency of 100 Hz and field of 1 Oe. Electrical-resistivity measurements were conducted
using a standard four-probe method in a PPMS-14. Pulsed-field experiments were performed up to 62 T using a magnet with 150 ms pulse duration and data were obtained via a fast data acquisition system operating with AC current in the kHz range. Contacts were made on freshly cleaved surfaces inside a glove box.

\begin{figure}[tbp]
\centering
\includegraphics[width=0.99\columnwidth]{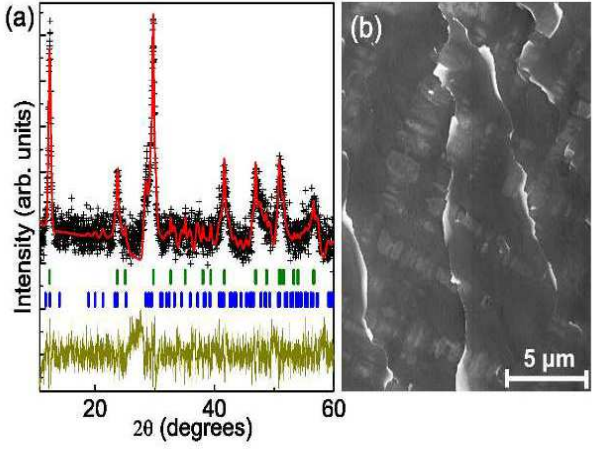} \vspace*{-0.5cm}
\caption{\label{fig1}(Color online) (a) Powder XRD pattern of K$_{0.50(1)}$Na$_{0.24(4)}$Fe$_{1.52(3)}$Se$_{2}$. The plot shows the
observed (+) and calculated (solid red line) powder pattern with the
difference curve underneath. Vertical tick marks represent Bragg
reflections in the I4/mmm (upper green marks) and I4/m (lower blue
marks) space group. (b) SEM image of the crystal.}
\end{figure}

The powder XRD data (Fig.\ \ref{fig1}(a)) demonstrate the phase purity
of our samples without any extrinsic peak present. The pattern is refined
in the space groups $I4/mmm$ and $I4/m$ with fitted lattice parameters
$a = 0.3870(2)$ nm, $c = 1.4160(2)$ nm and $a = 0.8833(2)$ nm, $c = 1.4075(2)$
nm, respectively, reflecting phase separation and small sample yield.\cite{Lazarevic,Ryan,Wang,Liu,Ricci,Li} With Na
substitution, the lattice parameter $a$ decreases while $c$ increases
when compared to K$_{0.8}$Fe$_{2}$Se$_{2}$, consistent with lattice parameters
of NaFe$_{2}$Se$_{2}$.\cite{Guo,YingTP} The average stoichiometry was
determined by EDX, measuring multiple positions on the crystal. The obtained composition
K$_{0.50(1)}$Na$_{0.24(4)}$Fe$_{1.52(3)}$Se$_{2.00(5)}$ suggests vacancies
on both K and Fe sites. FeSe-122 superconductors feature an intrinsic phase
separation into magnetic insulating and superconducting regions.\cite{Ryan,Wang,Liu,Ricci,Li} As shown in the SEM image of Fig.\ \ref{fig1}(b),
K$_{0.50(1)}$Na$_{0.24(4)}$Fe$_{1.52(3)}$Se$_{2}$ also exhibits a similar array
of superconducting grains in an insulating matrix. The observed pattern is
somewhat inhomogeneous [Fig.\ \ref{fig1}(b)] with sizes ranging from about
several microns to probably several tens of nanometers,\cite{WangZ} below
our resolution limit. It would be of interest to investigate the local structure and electronic properties of K$_{0.50(1)}$Na$_{0.24(4)}$Fe$_{1.52(3)}$Se$_{2.00(5)}$ since Na substitution provides chemically induced pressure which might suppress the phase separation similar as for Rb$_{1-x}$Fe$_{2-y}$Se$_{2}$.\cite{Bendele}

\begin{figure}[tbp]
\centering
\includegraphics[width=0.99\columnwidth]{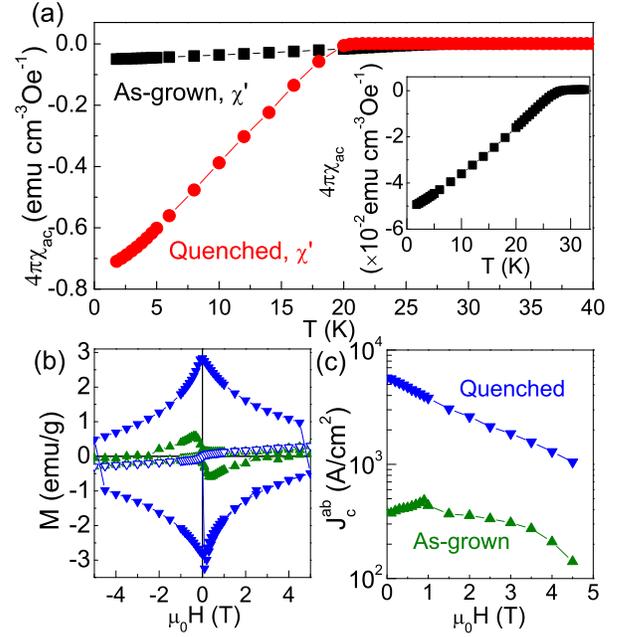} \vspace*{-0.5cm}
\caption{\label{fig2}(Color online) (a) Temperature dependence of
the ac magnetic susceptibilities of as-grown (magnified in the inset)
and quenched K$_{0.50(1)}$Na$_{0.24(4)}$Fe$_{1.52(3)}$Se$_{2}$. (b) Magnetic hysteresis loops of as-grown (triangles) and quenched
(inverted triangles) samples at $T = 1.8$ K (closed symbols) and
$T = 300$ K (open symbols) for $H\|c$. (c) Superconducting critical current
densities, $J_{c}^{ab}(\mu_{0}H)$, at $T = 1.8$ K.}
\end{figure}

The investigated single crystal becomes superconducting at
20 K after and at 28 K before the annealing and quenching procedure
[Fig.\ \ref{fig2}(a) main part and inset, respectively].\cite{Lei,Ryu}
For the quenched crystal, the superconducting volume fraction at 1.8 K increases
significantly up to 72\%, albeit with a reduction of $T_{c}$. The post-annealing and quenching process results in a
surface oxidation of some crystals which then dominates the magnetization
signal. However, Fe$_{3}$O$_{4}$ is not visible in either of our
laboratory or synchrotron X-ray studies.\cite{Lei,Ryu,HanF} The
magnetic hysteresis loops (MHL) of the quenched
K$_{0.50(1)}$Na$_{0.24(4)}$Fe$_{1.52(3)}$Se$_{2}$ single crystal reflects
the improvement in crystalline homogeneity since it is
much larger and symmetric when compared to an as-grown sample
[Fig.\ \ref{fig2}(b)] due to stronger pinning forces and bulk pinning.\cite{Lei}
Also similar to K$_{x}$Fe$_{2-y}$Se$_{2}$, there is an enhancement of the
in-plane critical-current density calculated from the Bean model:\cite{Bean,Gyorgy} $J_{c}^{ab}(\mu_{0}H)=\frac{20\Delta M(\mu_{0}H)}{a(1-a/3b)}$,
where $a$, $b$, and $c$ are the lengths of a rectangularly shaped crystal
($b > a > c$). In view of the improved volume fraction and homogeneity,
further investigations of the electronic transport
properties were performed on the quenched crystal.

\begin{table}[b]
\caption{Superconducting parameters of the quenched
K$_{0.50(1)}$Na$_{0.24(4)}$Fe$_{1.52(3)}$Se$_{2}$ single crystal.}
\begin{ruledtabular}
\begin{tabular}{ccccc}
 &$T_{c}$ & $(d\mu_{0}H_{c2}/dT)\vert_{T=T_{c}}$ & $\mu_{0}H_{c2}(0)$ & $\xi(0)$  \\
 & (K) &(T/K) & (T) & (nm) \\
\hline
H$\bot$c & 14.1(5) & -4.3(3) & 150$\sim$160 & 2.62$\sim$2.95 \\
H$\|$c & 14.1(5) & -1.1(2) & 38$\sim$48 & 0.75$\sim$0.79 \\
\end{tabular}
\end{ruledtabular}
\end{table}

The resistance of an inhomogeneous sample contains contributions from
both metallic ($R_{m}$) and nonmetallic ($R_{i}$) regions. At $T<T_{c}$,
due to superconductivity ($R_{m}=0$) the insulating part of the sample
is short-circuited. The insulating regions have a several orders of
magnitude higher resistivity than the metallic part;\cite{Shoemaker} hence, around $T_{c}$ and when $T\rightarrow0$ in the
high-field normal state $R(T)\approx R_{m}(T)$. This is similar to the
resistance of a polycrystalline sample in the presence of grain boundaries
and in agreement with the observation that insulating regions do not
contribute to the spectral weight in angular resolved photoemission data
in the energy range near $E_{F}$.\cite{YiM} In what follows below, we
focus on the temperature-dependent sample resistance, $R(T)$.

\begin{figure}[tbp]
\centering
\includegraphics[width=0.99\columnwidth]{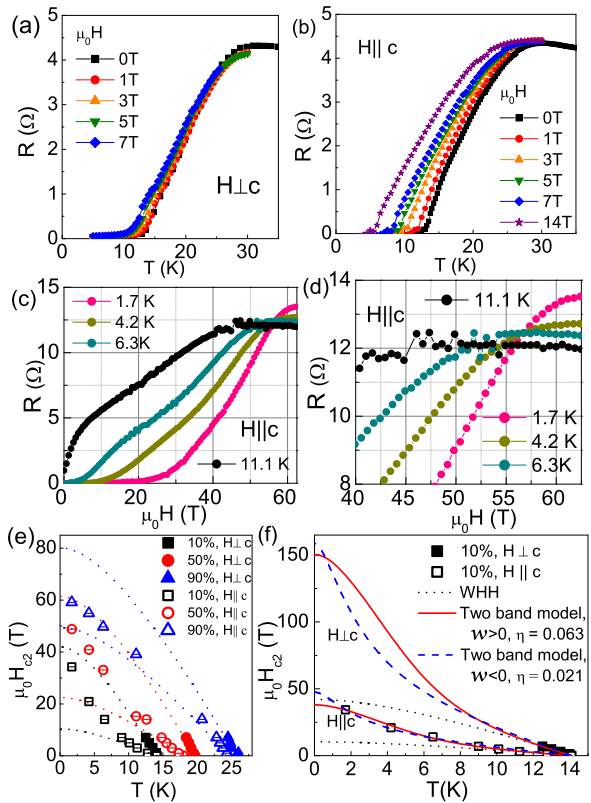} \vspace*{-0.5cm}
\caption{\label{fig3}(Color online) In-plane resistivity, $R_{ab}(T)$, of
K$_{0.50(1)}$Na$_{0.24(4)}$Fe$_{1.52(3)}$Se$_{2}$ for (a) $H\perp c$ and (b)
$H\| c$. $R_{ab}(T)$ measured at various temperatures
in pulsed magnetic fields up to 63 T in the full field range (c) and near the $H_{c2}(T)$ . (e) Temperature
dependence of the resistive upper critical field, $\mu_{0}H_{c2}$, determined
using three different criteria (10\%, 50\%, and 90\% of the normal-state value). Dotted lines are the WHH plots.
(f) Superconducting upper critical fields for $H\perp c$ (closed symbols) and
$H\| c$ (open symbols) using Eq.\ (1) with different pair breaking mechanisms:
(1) WHH (dotted line), (2) two-band model with {\scriptsize$\mathcal{W}$} $> 0$,
$\eta$ = 0.063 (solid line), and (3) two-band model with
{\scriptsize$\mathcal{W}$} $< 0$, $\eta$ = 0.021 (dashed line).}
\end{figure}

The superconducting transition in $R_{ab}(T)$ is rather wide and shifts
to lower temperatures in applied magnetic fields [Figs.\ \ref{fig3}(a-d). The shift is more pronounced for $H\|c$, which implies an anisotropic
$\mu_{0}H_{c2}$. The temperature-dependent upper critical fields shown in
Fig.\ \ref{fig3}(c) were determined from the resistivity drops to 90\%, 50\%,
and 10\% of the normal-state value. It is clear that all experimental data feature a similar temperature dependence irrespective of the criteria used. All data for $H\|c$ are above the expected values for the single band Werthamer-Helfand-Hohenberg (WHH) model (dotted lines). We proceed our further analysis using the 10\% values, similar as done for LaFeAsO$_{0.89}$F$_{0.11}$.\cite{Hunte} The $H_{c2}(T)$ curves are linear for $H\perp c$ near $T_{c}$
and show an upturn at low T for $H\| c$ [Fig.\ \ref{fig3}(e)]. The initial
slope near $T_{c}$ for $H\perp c$ is much larger than for $H\| c$
[Fig.\ \ref{fig3}(f) and Table I]. These slopes are similar to values for
as-grown and quenched K$_{x}$Fe$_{2-y}$Se$_{2}$.\cite{Mun,Lei1}

There are two basics mechanisms of Cooper-pair breaking by magnetic field
in a superconductor. Orbital pair breaking imposes an orbital limit due to
the induced screening currents, whereas the Zeeman effect
contributes to the Pauli paramagnetic limit of $H_{c2}$. In the single-band
WHH approach, the orbital critical field is
given by $\mu_{0}H_{c2}(0)$ = -0.693$T_c(d\mu_{0}H_{c2}/dT)
|_{T=T_{c}}$.\cite{WHH} For K$_{0.50(1)}$Na$_{0.24(4)}$Fe$_{1.52(3)}$Se$_{2}$, this leads to 42(3) T for $H\perp c$ and 10(2) T for $H\| c$
[Fig.\ \ref{fig3}(f)]. On the other hand, the Pauli-limiting field is given by
$\mu_{0}H_{p}(0)=1.86T_{c}(1+\lambda_{e-ph})^{1/2}$, where $\lambda_{e-ph}$ is
the electron-phonon coupling parameter.\cite{Orlando} Assuming
$\lambda_{e-ph}$ = 0.5, which is a typical value for a weak-coupling BCS
superconductor,\cite{Allen} $\mu_{0}H_{p}(0)$ is 32(1) T. This is larger
than the orbital pair-breaking field for $H\| c$ estimated above, yet
smaller than the value for $H\perp c$, which possibly implies that
electron-phonon coupling is much stronger than for typical weak-coupling
BCS superconductors.

The experimental data for $\mu_{0}H_{c2}(0)$ lie above the expected
values from WHH theory [Fig.\ \ref{fig3}(f)], suggesting that multiband
effects are not negligible. In the dirty limit, the upper critical
field found for the two-band BCS model with orbital
pair breaking and negligible interband scattering is:\cite{Gurevich}
\begin{eqnarray}
&& a_0 \left[ \ln t + U(h) \right]\left[ \ln t + U(\eta h) \right] +
a_2 \left[ \ln t + U(\eta h) \right] \nonumber \\
&&\quad \quad \quad \quad \quad \quad \quad \quad \quad \quad
+a_1 \left[ \ln t + U(h) \right] = 0, \\
&& U(x) = \psi(1/2+x) - \psi(1/2),
\end{eqnarray}
where $t=T/T_{c}$, $\psi(x)$ is the digamma function, $\eta=D_2/D_1$,
$D_1$ and $D_2$ are diffusivities in band 1 and band 2, $h=H_{c2}D_{1}/
(2\phi_{0}T)$, and $\phi_{0}=2.07\times10^{-15}$ Wb is the magnetic flux quantum. $a_{0}=2${\scriptsize$\mathcal{W}$}$/\lambda_{0}$, $a_{1}=1+\lambda_{-}/
\lambda_{0}$, and $a_{2}=1-\lambda_{-}/\lambda_{0}$, where,
{\scriptsize$\mathcal{W}$}$=\lambda_{11}\lambda_{22}-\lambda_{12}\lambda_{21}$,
$\lambda_{0}=(\lambda_{-}^{2}+4\lambda_{12}\lambda_{21})^{1/2}$, and
$\lambda_{-}=\lambda_{11}-\lambda_{22}$. $\lambda_{11}$ and $\lambda_{22}$
are pairing (intraband coupling) constants in band 1 and 2, and
$\lambda_{12}$ and $\lambda_{21}$ quantify interband couplings between
band 1 and 2. For $D_{1}=D_{2}$, Eq. (1) simplifies to the one-band model
(WHH) in the dirty limit.\cite{WHH} When describing our data by use of the two-band BCS model fitting, we consider two different cases,
{\scriptsize$\mathcal{W}$} $> 0$ and {\scriptsize$\mathcal{W}$} $< 0$,
which imply either dominant intraband or dominant interband coupling,
respectively. The solid lines in Fig.\ \ref{fig3}(f) are fits using
Eq. (1) for $\lambda_{11}=\lambda_{22}=0.5$ and $\lambda_{12}=
\lambda_{21}=0.25$ which indicates strong intraband coupling.\cite{Jaroszynski,LeiH2} The
extrapolated $\mu_{0}H_{c2}(0)$ is $\sim$38 T for
$H\| c$ and $\sim$150 T for $H\perp c$. Further, the dashed lines in
Fig.\ \ref{fig3}(f) show fits with $\lambda_{11}=\lambda_{22}=0.49$
and $\lambda_{12}=\lambda_{21}=0.5$ for strong interband coupling\cite{Jaroszynski,LeiH2} that
give $\mu_{0}H_{c2}(0)$ $\sim$48 T for $H\| c$ and $\sim$160 T for $H\perp c$.

From these fits we obtain $\eta$ values of 0.063
and 0.021 for dominant intraband ({\scriptsize$\mathcal{W}$} $>$ 0) and
interband ({\scriptsize$\mathcal{W}$} $<$ 0) coupling, respectively,
i.e., largely different $D_1$ and $D_2$ implying
different electron mobilities in the two bands. The upward curvature of
$\mu_{0}H_{c2}(T)$ is governed by $\eta$; it is more
pronounced for $\eta\ll 1$. The large difference in the intraband
diffusivities could be due to pronounced differences in effective masses,
scattering, or strong magnetic excitations.\cite{Gurevich,Jaroszynski}
The fit results are not very sensitive to the choice of the coupling
constants, yet they mostly depend on $\eta$. This indicates either
similar interband and intraband coupling strengths or that their
difference is beyond our resolution limit. Our results are consistent with the data obtained on pure crystals, i.e., the large difference of electronic diffusivities for different Fermi surface sheets is maintained in the doped crystal.\cite{Gasparov} This is in agreement with the band-structure calculations that showed negligible contribution of K to the Fermi surface and density of states at the Fermi level.\cite{Kreisel,Yan} On the other hand, we find no enhancement of the superconducting $T_{c}$ with Na substitution in K$_{x}$Fe$_{2-y}$Se$_{2}$ ($T_{c}$ $\sim$ 30 K). This is somewhat surprising because Na$_{x}$Fe$_{2}$Se$_{2}$ and NaFe$_{2}$Se$_{2}$ that crystallize in \textit{I4/mmm} space group have $T_{c}$'s of 45 and 46 K.\cite{YingTP} The Na substitution might affect the magnetic order in phase separated K$_{x}$Fe$_{2-y}$Se$_{2}$ since the existence of a large magnetic moment in the antiferromagnetic phase was proposed to be important for the relatively high $T_{c}$'s.\cite{HuangMS}

Due to the limited data points, it is difficult to unambiguously
estimate $\mu_{0}H_{c2}(0)$ for $H\perp c$. Based on results reported
for similar Fe-based superconductors, NdFeAsO$_{0.7}$F$_{0.3}$,\cite{Jaroszynski}
(Ba,K)Fe$_{2}$As$_{2}$,\cite{Yuan} and K$_{0.8}$Fe$_{1.76}$Se$_{2}$,\cite{Mun} $\mu_{0}H_{c2}(0)$ shows a pronounced
upward curvature for $H\| c$ while it tends
to saturate for $H\perp c$. The real $\mu_{0}H_{c2}(0)$ for $H\perp c$
might be smaller than we estimated. The calculated coherence lengths, using
$\mu_{0}H_{c2}^{\perp}(0)=\phi_{0}/2\pi\xi_{\perp}(0)\xi_{\|}(0)$ and
$\mu_{0}H_{c2}^{\|}(0)=\phi_{0}/2\pi\xi_{\perp}(0)^{2}$ based on the
two-band BCS fit results, are similar to values obtained for as-grown
and quenched K$_{x}$Fe$_{2-y}$Se$_{2}$ and are shown in Table I.\cite{Mun,Lei1}

\begin{figure}[tbp]
\centering
\includegraphics[width=0.95\columnwidth]{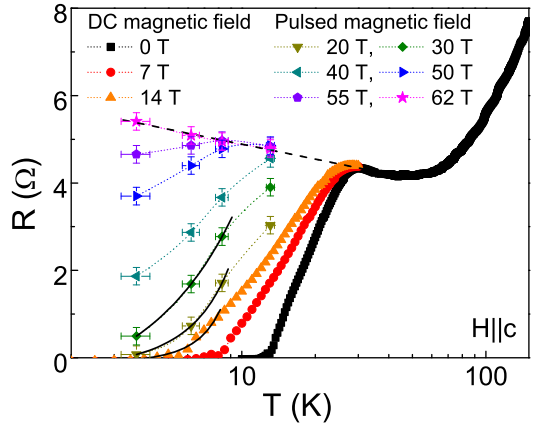}
\caption{\label{fig4}(Color online) Temperature dependence of the
resistance in several DC and pulsed magnetic fields for $H\| c$.}
\end{figure}

Superconductivity in K$_{0.50(1)}$Na$_{0.24(4)}$Fe$_{1.52(3)}$Se$_{2}$
is completely suppressed above about 60 T for $H\| c$,
allowing for a clear insight into the low-temperature electronic transport in the normal state (Fig.\ \ref{fig4}). Interestingly,
we do not observe metallic transport below about 40 K implying that a superconductor-to-insulator transition (SIT) is induced
in high magnetic fields. Kondo-type magnetic scattering is not very
likely since a field of 62 T should suppress spin-flip scattering.\cite{Ando}
A thermally activated semiconductor-like transport or
variable range hopping (VRH) as occurring for Anderson
localization is unlikely since the resistance in 62 T cannot be fit by
$\ln R \propto -1/T$, $\ln R \propto T^{-\beta}$, with $\beta$ = 1/2,
1/3, 1/4, and $\ln R \propto \ln T$.\cite{Mott,Abrahams,Gorkov}
Instead, the resistance increases logarithmically with decreasing
temperature in the normal state at 62 T as shown with the dashed line
in Fig.\ \ref{fig4}. Hence, the SIT might originate from the granular
nature of K$_{0.50(1)}$Na$_{0.24(4)}$Fe$_{1.52(3)}$Se$_{2}$. In a
bosonic SIT scenario, Cooper pairs are localized in granules.\cite{Gantmakher,BeloborodovRMP} When $H > H_{c2}$, virtual Cooper
pairs form, yet they cannot hop to other granules when $T\rightarrow 0$
which induces the increase in resistivity as temperature decreases.
The grain size can be estimated from $H_{c2}^{0}\sim \phi_{0}/L\xi$,
where $L$ is the average grain radius and $\xi \approx 0.77$ nm is the
average in-plane coherence length. The obtained $L=62$ nm is in
agreement with the phase-separation distance. The bosonic SIT mechanism
in granular superconductors predicts $R=R_{0}\exp(T/T_0)$ (`inverse
Arrhenius law') in the superconducting region near the SIT when
$H < H_{c2}$ due to the destruction of quasi-localized Cooper pairs
by superconducting fluctuations. Our data in 14, 20, and 30 T might be
fitted with this formula (solid lines in Fig.\ \ref{fig4}).
We note that $R(H)$ near $\mu_{0}H_{c2}(0)$ is non-monotonic, similar as for granular Al and La$_{2-x}$Sr$_{x}$CuO$_{4}$.\cite{Ando,BeloborodovRMP,Gerber}

In summary, we reported the multiband nature of superconductivity
in K$_{0.50}$Na$_{0.24}$Fe$_{1.52}$Se$_{2}$ as evidenced in
the temperature dependence of the upper critical field and a SIT in
high magnetic fields. Granular type-I but also copper-oxide
superconductors are also intrinsically phase separated on the
nanoscale.\cite{Strongin,Imry,Lang,Zeljkovic} Hence, a SIT in high
magnetic fields seems to be connected with the
intrinsic materials' granularity in inhomogeneous superconductors.
This suggests that the insulating states found in cuprates as a
function of magnetic field\cite{Ando,Boebinger} or doping\cite{Fukuzumi}
might involve Josephson coupling of nanoscale grains as opposed to
quasi-one-dimensional metallic stripes bridged by Mott-insulating
regions in the spin-charge separated picture.\cite{Emery}

\begin{acknowledgments}

Work at Brookhaven is supported by the U.S.\ DOE under Contract
No.\ DE-AC02-98CH10886 and in part by the Center for Emergent
Superconductivity, an Energy Frontier Research
Center funded by the U.S.\ DOE, Office for Basic Energy Science
(C.P.). We acknowledge the support of the HLD at HZDR, member of the European Magnet Field Laboratory (EMFL). CP acknowledges support by the Alexander von Humboldt
Foundation.

\end{acknowledgments}

\ddag\ hryu@bnlgov and petrovic@bnl.gov
\dag Present address: European XFEL GmbH, Notkestrasse 85, 22607 Hamburg, Germany

\end{document}